\documentclass[prl,aps,showpacs,superscriptaddress,twocolumn ]{revtex4}

\usepackage{graphicx}
\usepackage{amsmath}
\usepackage{amsfonts}
\usepackage{amssymb}
\usepackage{epsf}
\usepackage[usenames,dvipsnames]{color}
\usepackage{hyperref}
\newcommand{\beq}{\begin{equation}}
\newcommand{\eeq}{\end{equation}}
\def\beqa#1\eeqa{\begin{align}#1\end{align}}
\def\bcorr #1 \ecorr{ {\color{red} #1 } }
\newcommand{\bu}{ \mathbf u }
\newcommand{\bp}{ \mathbf p }
\newcommand{\br}{ \mathbf r }
\newcommand{\er}{ \mathbf e_r }

\DeclareMathOperator*{\dt}{\partial_t }

\DeclareMathOperator{\bP}{\bf p}
\DeclareMathOperator{\bI}{\bf I}

\begin{document}
\title{Blast dynamics in a dissipative gas}

\author{M. Barbier}
\affiliation{Department of Ecology and Evolutionary Biology, Princeton University, Princeton, NJ 08544, USA}

\author{D. Villamaina}
\affiliation{Laboratoire de Physique Th\'eorique de l'ENS (CNRS UMR 8549) and Institut de Physique
Th\'eorique Philippe Meyer, 24 rue Lhomond 75005 Paris - France}

\author{E. Trizac}
\affiliation{Laboratoire de Physique Th\'eorique et Mod\`eles Statistiques (CNRS UMR 8626), B\^atiment 100, Universit\'e Paris-Sud, 91405 Orsay cedex, France}

\date{\today}

\begin{abstract}
The blast caused by an intense explosion has been extensively studied in conservative fluids, where the Taylor-von Neumann-Sedov hydrodynamic solution is a prototypical example of self-similarity 
driven by conservation laws. In dissipative media however, energy conservation is violated, yet a distinctive self-similar solution appears. It hinges on the decoupling of random and coherent motion 
permitted by a broad class of dissipative mechanisms. This enforces a peculiar layered structure in the shock, for which we derive the full hydrodynamic solution, validated by a microscopic 
approach based on Molecular Dynamics simulations. 
We predict and evidence a succession of temporal regimes, as well as a long-time corrugation instability, also self-similar, which disrupts the blast boundary. These generic results may apply 
from astrophysical systems to granular gases, and invite further cross-fertilization between microscopic and hydrodynamic approaches of shockwaves.
\end{abstract}


\pacs{45.50.-j,45.70.-n,47.40.Rs}

\maketitle

A blastwave follows the rapid and localized release of a large amount of energy in a medium. The physics 
community  got seasonably interested in the dynamics of such shocks in air in the early 1940s.
Taylor \cite{Taylor}, von Neumann \cite{vonNeumann} and Sedov \cite{Sedov} independently understood that, as a result of the global conservation of mass and energy, 
the extension $R$ of the 
blast had to grow with time like a power law $t^{\delta}$, with $\delta=2/5$ (or $2/(d+2)$ 
in dimension $d$)\cite{argument}. From a few publicly available snapshots of the 
blast at different times, Taylor could estimate within 10\% the 
strength of the Trinity detonation in 1945, at the time a classified information \cite{Bare96}.

Remarkably, the hydrodynamic description of the flow inside the blast, now  known as the Taylor-von Neumann-Sedov solution (TvNS), 
is self-similar in time, depending only on the rescaled radial distance $r/R(t)$.  This similarity is \textit{of the first kind} \cite{Bare96}
i.e. driven by global invariants, and all exponents can be derived by dimensional analysis. 
This solution  found widespread relevance beyond its initial realm, notably in plasma physics  to describe
laser-induced shocks~\cite{Edens,Moore} and in astrophysics for the
evolution of supernova remnants~\cite{CMB}. 
However, it proves essential for a wealth of applications to relax some of the conservation laws~\cite{OsMK88,Bare96}, especially allowing for energy 
production or dissipation: on the shock boundary (e.g. a chemical reaction front) or in the bulk (e.g. collisional or radiative losses). This is usually expected 
to entail self-similarity \textit{of the second kind}, where scaling exponents are no longer globally fixed, but depend continuously on parameters of the dynamics~\cite{Bare96}. 

\begin{figure}[htb]
\centerline{\includegraphics*[width=7cm,clip=true]{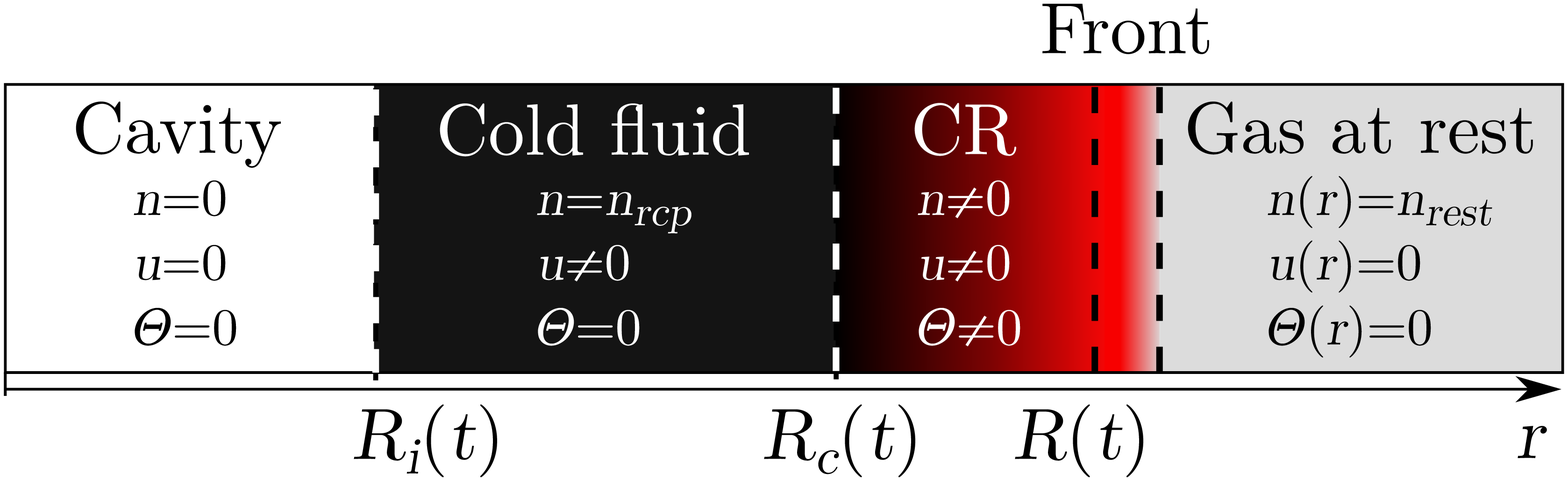}}
\centerline{\hspace*{6pt}\includegraphics*[width=7.25cm,clip=true]{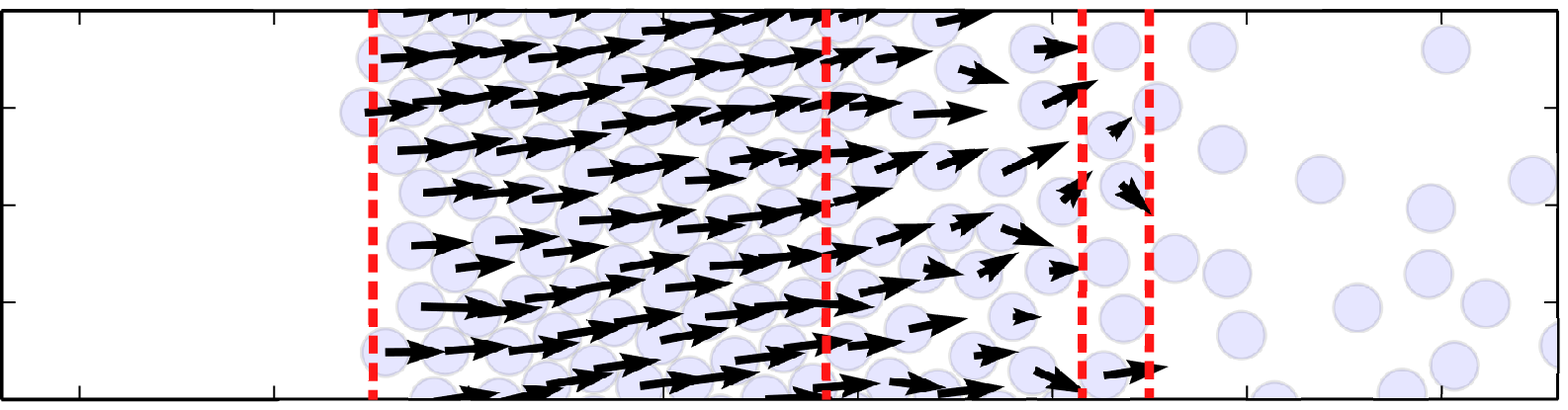}}
\caption{Section of the blastwave: cartoon (top) and hard sphere MD simulation (bottom). Dissipation causes particles to accrete into a dense, hollow shell. 
Eqs.\eqref{eq:coolingprofiles}-\eqref{eq:coldprofiles} reveal its layered structure. First, the shock front where particles from the blast collide and mix with those at rest, creating incoherent motion. Then, a fixed-width cooling region (CR) where further compression occurs as temperature is dissipated, and finally a self-similar cold fluid region where velocities are well aligned, around a central cavity. This stands in stark contrast with the elastic case, where the bulk of the blast is comprised of a single, entirely self-similar region.}
\label{Fig:structure}
\end{figure}

The subject of our inquiry, a blast with bulk energy dissipation, runs contrary to that expectation. Understanding its similarity properties requires connecting two levels of analysis, coarse-grained and microscopic, that have remained largely impervious to each other. 
The abundant literature following  the TvNS model  has focused on  global scaling laws or hydrodynamic models.
Meanwhile, the study of granular gases~\cite{BrPo04} provides 
prototypical dissipative media where  experiments~\cite{BoCK09,BoKe13} and particle-based simulations~\cite{Jabeen1}
have been performed, but a continuum view is missing.
It is our purpose here to bridge this gap~\cite{rque10}. We provide an analytical understanding for three numerical observations: the
spatial profiles for hydrodynamic fields, the scaling regimes exhibited by $R(t)$, and a previously unreported corrugation
instability that distorts the shockwave at late times. Unraveling the wave structure 
is key in explaining these properties; 
Fig. \ref{Fig:structure} summarizes its main features, with a dense shell divided in three regions. 
 The corrugation instability, also self-similar, is unique to the dissipative blast and
differs significantly from those described in various other blastwaves \cite{ViRy87and89,Kushnir05}.
We will argue that, under broad assumptions, our results are largely independent of the mechanism at play in energy dissipation.

\paragraph*{Model and previous results:}  Our model system is a granular gas of identical spherical grains with radius $\sigma$ and unit mass, 
where inelastic binary collisions conserve momentum but dissipate kinetic energy. Core results will not depend on specific dissipation mechanisms, hence we opt for simplicity: energy loss is
quantified by a fixed restitution coefficient $0 \leq \alpha\leq 1$ \cite{BrPo04}; encounters are dissipative 
when $\alpha<1$ and energy conserving (elastic) for $\alpha=1$. 
We set a break-shot initial condition, where all grains are at rest
except within a small region \cite{AnKR08,Jabeen1}.
A cascade of collisions follows, with an ever growing number of particles in motion,
which forms the blast as observed in Fig.~\ref{Fig:structure}. We define its radius $R(t)$ as the distance from the center to the innermost particle at rest, and the shock front as all moving particles that count grains at rest among their nearest neighbors.
Since the external medium is motionless, strong shock (infinite Mach number) conditions are ensured for any initial energy release~\cite{rqene}. {\it A priori}, the dynamics are specified by two parameters: $\alpha$ and the volume fraction $\phi_{\text{rest}} = n_{\text{rest}} \mathcal V_d(\sigma) $, where $n_{\text{rest}}$ is the number density of particles in the gas at rest, and $\mathcal V_d(\sigma)$ is the volume of a grain in dimension $d$.
In the remainder, numerical results are obtained from Molecular Dynamics (MD) simulations \cite{BrPo04,rque}.

The first property of interest is the scaling law(s) obeyed by the radius $R(t)$. As the TvNS scaling crucially hinges on energy conservation, 
relaxing the latter generically causes self-similarity to break down or cross over to the second kind -- it is then 
sensitive to microscopic parameters 
and derivable through methods such as renormalization, but not dimensional analysis \cite{Bare96}. However, following an argument due to Oort \cite{Oort51} 
for astrophysical systems, a blast in a gas with \textit{bulk} dissipation should tend to form a thin, hollow shell that slows down only by accreting more material. 
Its total radial momentum, of order 
$R^d dR/dt$, is thus constant, at odds with the energy-conserving case (as explained below). This implies $R \propto t^\delta$ with 
$\delta = 1/(d+1)$, or $1/4$ in three dimensions -- smaller than its conservative counterpart $2/(d+2)$. This solution is known as the Momentum Conserving Snowplow (MCS)~\cite{CMB} and is self-similar of the first kind. 
Recent numerical studies of granular gases have confirmed both this scaling law and the hollow structure of the blast for any $\alpha<1$, 
although the shell is thick due to the  high densities considered~\cite{Jabeen1}. 

\paragraph*{Hydrodynamic description:}
Previous works on the granular blast have stopped short of investigating its spatial structure beyond these simple arguments. We turn to a continuum 
description, which will shed light on the peculiar shell in Oort's argument, and reveal its long-term instability.
To establish a closed set of hydrodynamics equations, we define the \textit{granular
temperature} (or energy of random motion)  of the medium through the variance of local velocity
fluctuations \cite{BrPo04,BDKS98,Gold03,rqtmp}. The coupling between this temperature field $\Theta(\br)$ and the density and
velocity fields $n(\br)$ and ${\bf u}(\br)$ is derived in the dense
fluid transport framework for inelastic hard spheres
\cite{GaDu99}, which generalizes earlier descriptions of dilute
systems \cite{BDKS98,JeSa83}  
\begin{subequations}
\label{eq:hydro}
\beqa & \dt n + \nabla (n\bu) = 0  \\
& (\dt + \bu.\nabla) \bu + \dfrac 1 n \nabla.\bP =0  \label{eq:uhydro}\\
&  n(\dt + \bu.\nabla) \Theta + \frac{2}{d} (\bP\cdot\nabla)\cdot\bu  =-\Lambda  .
\label{eq:Ehydro}
\eeqa
\end{subequations}
The energy sink term takes the form \cite{GoSh95,GaDu99} 
\beq 
\Lambda \,=\, \omega (1-\alpha^2) n \Theta
\label{eq:dissip}
\eeq
and $\omega = \omega_0 n\sigma^{d-1}\Theta^{1/2}$ is the local collision frequency, proportional to the average relative velocity and the inverse mean free path, with $\omega_0$ a dimensionless constant.
These equations are closed by specifying the pressure tensor ${\bP}$ with a constitutive relation, discussed below, which is safely taken of zeroth order in gradients, neglecting heat conduction~\cite{note2}.

In the particular elastic case ($\alpha=1$, hence $\Lambda=0$) with isotropic pressure $\bP=p \bI$ (${\bI}$ being the identity matrix), the Euler equations for a perfect fluid are 
 recovered. Within the blast, the fields then assume a scaling form
\beqa
n(\br,t) \,=\, n_{\text{rest}} \, M(\lambda), \hspace{10pt} 
\bu(\br,t)\,=\, \dfrac{\br}{t} \, V(\lambda) \nonumber\\ 
\Theta(\br,t)\,=\, \dfrac{r^2}{t^2} \, T(\lambda), 
\hspace{10pt} 
p(\br,t) \,=\, n_{\text{rest}} \,\dfrac{r^2}{t^2}P(\lambda) 
\label{eq:scaling_elastic}
\eeqa
where $\br=r \,\er$ denotes the position relative to the center of the blast with $r=|\br|$, and $\lambda = r/R(t)$ is the scaling variable. The profiles $M$, $V$, $T$ and $P$ defined by Eq.~\eqref{eq:scaling_elastic} are dimensionless, isotropic and time-independent, and can be evidenced throughout the self-similar blast expansion by appropriately rescaling the density, velocity, temperature and pressure fields.
With the further assumption of an ideal gas constitutive 
relation $p = n \,\Theta$, 
Eqs. \eqref{eq:hydro} and \eqref{eq:scaling_elastic}  together admit the classic TvNS solution \cite{Taylor,vonNeumann,Sedov,Bare96}.
It is noteworthy that there is a unique velocity scale in the elastic problem: 
 $\bu$ and $\sqrt\Theta$  exhibit the same scaling in $R(t)/t$, meaning that coherent and incoherent
motion remain coupled.
This solution exhibits a simple structure,
with a boundary layer of fixed size (the shock front) around an isotropic, self-similar bulk. 
The front, where discontinuities in the hydrodynamic fields occur, can be
defined microscopically as the thin mixed region where mobile and immobile particles coexist.

\begin{figure}[htb]
\centerline{\includegraphics*[width=9 cm]{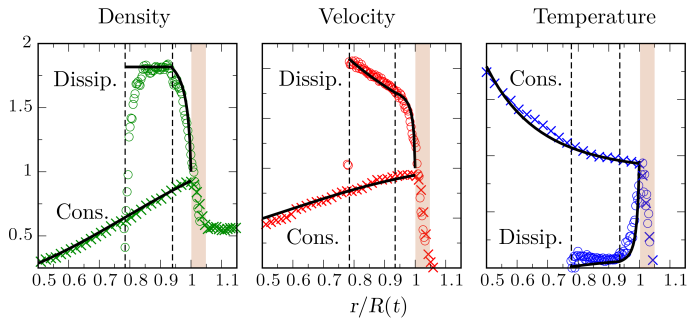}}
\caption{ Hydrodynamic profiles from MD simulations for dense conservative (crosses) and dissipative (circles) fluids with $\phi_{\text{rest}} = 0.3$, and analytical 
solutions (solid lines). Profiles are rescaled by the Rankine-Hugoniot boundary value (see \cite{SuppMat}), from left to right: $n(\lambda)/n_{RH}$, $u(\lambda)/u_{RH}$ and 
$\Theta(\lambda)/\Theta_{RH}$ with 
$\lambda=r/R(t)$. The conservative solution is partitioned into the gas at rest, the shock front (shaded area) and the self-similar bulk. This is 
the first validation of a TvNS-like solution in a dense fluid.
For any dissipation i.e. $\alpha<1$ (here $\alpha=0.8$), the bulk structure becomes threefold, 
as shown by dashed vertical lines: the cooling region between the front and $R_c\approx 0.93\, R$; the maximally dense cold fluid down to 
$R_i\approx 0.78\, R$;  and the central cavity.  }
\label{Fig:profiles}
\end{figure}

\paragraph*{Results and discussion\/:} In the dissipative case, the front is unchanged, but the core becomes more complex, and cannot be described by fields with a simple scaling form. Indeed, the dissipation term in Eq. \eqref{eq:Ehydro} 
depends explicitly on an additional time scale, 
the collision time  $\omega^{-1}$, and is consequently incompatible with the scaling \eqref{eq:scaling_elastic}. 
In the front, the spread in velocity is of order $\dot R$, since particles advancing at that speed collide with others at rest. Incoherent motion is continuously generated by these collisions, so that at the boundary $\Theta \sim  \dot R^2(t)$, as in the elastic case. Inserting this ansatz in Eq. \eqref{eq:Ehydro}, the dissipation term grows in magnitude compared to transport terms by a factor $\omega R(t)/ \dot R(t) \sim R(t)$: asymptotically, temperature is dissipated too fast to be advected on distances comparable to $R(t)$. Dissipation instead acts over a distance $\dot R/\omega$ that is here time independent; this
creates a travelling wave-like zone behind the front, the `cooling region'. 
Then, moving further toward the interior of the blast, we find
a distinct cold layer  where temperature is negligible and particle velocities are aligned; dissipation is eliminated, and fields once again assume a self-similar form. 
This region reaches a density close to random close packing, echoing the clustering tendency of granular media~\cite{BrPo04}. 
Finally, as all the inner material accretes into this dense layer, an empty cavity opens at the center and the blastwave forms a hollow shell, in line with Oort's
argument~\cite{Oort51}.
The validation of this multilayered structure in MD simulations (see Fig.~\ref{Fig:structure} 
and \ref{Fig:profiles}) is a crucial result that drives all further analysis. We expect it to hold beyond our model system: 
the decoupling of length or energy scales that shapes the layers only requires $\omega$ to increase with $n$ and $\Theta$.

By contrast with past attempts at elucidating the blast's structure~\cite{Bert86}, we argue that a consistent description requires a dense fluid equation of state, to account for compression in the shell. 
We find a solution to equations~\eqref{eq:hydro}
with an anisotropic pressure tensor acting only along the radial direction $\er$
\begin{equation}
\bp \,=\, n \, \Theta \, Z(n) \, \er \otimes \er 
\label{eq:pression_aniso}
\end{equation}
where $Z(n)$ is a diverging function of density, 
accounting for steric effects \cite{GoSh95,GaDu99,SuppMat}. Such strongly anisotropic pressure is frequently observed in granular systems \cite{anisotropic}, and the divergence in $Z(n)$ allows the cold region, 
despite its vanishing temperature, to exhibit a finite pressure field. This is essential for the existence of self-similar profiles in analogy with Eqs.~\eqref{eq:scaling_elastic}. 
Boundary conditions on these profiles are set by a fixed-width layer, comprised of the shock front where temperature is created, and the cooling region
where it is dissipated. We describe this layer in a flux-difference form \cite{GoSh95}, which  generalizes Rankine-Hugoniot conditions~\cite{Landau} to the dissipative case and over a finite distance \cite{SuppMat}:
in this thin layer the system is assumed to be quasi-1D, allowing us to simplify
Eqs. \eqref{eq:hydro} and integrate them between $R(t)$ and $r=R(t)-x$ giving 
\beq
n(x)=n_{\text{rest}} M(x), \hspace{5pt} 
u=\dot R \left[1-\frac{1}{M}\right],\hspace{5pt} 
p=n_{\text{rest}} \dot R \, u
\label{eq:coolingprofiles}
\eeq
where all three fields are parameterized by the compression $M(x)$ obeying the following equation (which may be integrated
numerically for any choice of $Z(n)$) 
\beqa
(M-1) &\left[\frac{d}{2} Z^{-1}+1\right]=
\nonumber \\ 
&\dfrac{M^2}{2} -\omega_0(1-\alpha^2) \int_{0^-}^x dx'
\left(\dfrac{d(M-1)}{2 Z} \right)^{3/2} .
\label{eq:Blast.Hyd.cooleq}
\eeqa
Higher-order transport terms neglected in Eqs.~\eqref{eq:hydro} may in fact 
intervene in this layer, which has no growing typical length scale; however,
this simplified analysis proves accurate enough.	
 
In the cold region, we
can assume scaling forms similar to Eqs.~\eqref{eq:scaling_elastic}, although temperature is vanishingly small.
An analytical solution is possible, as the fluid density is fairly approximated by its random close packing value $n_{\text{rcp}}=n_{\text{rest}}M_{\text{rcp}}$ corresponding to a volume fraction $\phi_{\text{rcp}} \approx 0.84$ for $d=2$ or $0.64$ for  $d=3$, hence
\beqa 
&V(\lambda)= \delta \left( 1 -M_{\text{rcp}}^{-1}\right) \lambda^{-d}, \nonumber \\ 
& P(\lambda)  = \delta^2\,\lambda^2\left( 1 -M_{\text{rcp}}^{-1}\right)\left(M_{\text{rcp}}(
\lambda^d -1) + 1 \right) .
\label{eq:coldprofiles} 
\eeqa
The profiles thus obtained are seen in Fig. \ref{Fig:profiles} to fare remarkably against MD simulations, despite the piecewise nature of the model. 
No fitted parameter is necessary.
This is a crisp validation of the hydrodynamic view developed here as, to the best of our knowledge, no other model for blasts (dissipative or conservative) has been 
successfully applied to a dense fluid or supported by microscopic simulations. 
As anticipated, the flow velocity is maximal on 
the inner boundary of the cold region: the shell is pushed outward by its innermost particles, while the outermost slow down with dissipation and accrete onto the incoming ``snowplow''. At odds with the conservative case, coherent flow thus decouples from thermal agitation; we now see that this is the cornerstone of the similarity solution.

From the tensorial form of pressure, Eq. (\ref{eq:pression_aniso}), it can be shown by 
integrating Eq.~\eqref{eq:uhydro} that 
radial momentum $\Pi = \int n u \,r^{d-1} \,dr$ within a given solid angle ($d=3$) or angular sector ($d=2$), is conserved.
This formally demonstrates the invariant suggested by previous authors \cite{Oort51,Jabeen1} which yields the growth exponent $\delta = 1/(d+1)$, as sketched above.
The reason for the conservation of $\Pi$ is twofold: first, the central pressure at $r=0$ vanishes in the cavity, and second, there are no orthoradial exchanges
of momentum, due to the decoupling of coherent (radial) and incoherent (partly orthoradial) motion.
In the elastic case however $(\alpha=1)$, the opposite statements hold, and the expansion is dominated by the central pressure,
which causes $\Pi$ to increase with time, leading to $\delta >  1/(d+1)$. On shorter timescales, the evolving interplay of these forces gives rise to a succession of intermediate regimes, such as the Pressure-Driven Snowplow~\cite{CMB}, which we confirm in MD simulations \cite{SuppMat}.

\begin{figure}[htb]
\includegraphics*[width=7.5cm,clip]{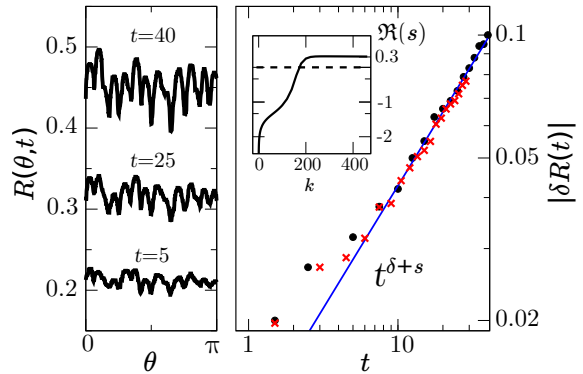}
\centerline{}

\caption{Left: Shock radius by angular sector $R(\theta,t)$ at successive
  times (see~\cite{SuppMat} for its definition). The line thickness represents the mean free path in the shock front, or about half its average width $w\approx 0.01$. Right: Corrugation width $\delta R \sim t^{\delta+s}$ with theoretical exponent $s\simeq 0.3$  (solid line) and numerical validation for $\alpha=0.3$ (circles) and $\alpha=0.8$ (crosses). Inset: Real part of the dispersion relation $s(k)$ for the unstable mode, crossing the marginal stability line $\Re(s)=0$ (dashed line), with a plateau at $\Re(s)\approx 0.3$. }
\label{Fig:cartoon}
\end{figure}

Finally, we evidence an instability arising at late times, which disrupts the MCS solution analyzed above and manifests as a growing corrugation of the shell 
(see Fig.~\ref{Fig:cartoon}). Numerics show
that this growth follows a power law, suggesting that
it stems from the self-similar cold region. While other
instabilities can be found in TvNS shockwaves under
specific conditions~\cite{Grun91}, we stress the generic character of
the instability discussed here. Necessary conditions are the  conservation of radial momentum and a vanishing pressure on the inner boundary of the shell, 
which explain the absence of this phenomenon in conservative gases, and distinguish it from other instabilities of granular systems~\cite{sirmas}. We perform a linear stability analysis, focusing here on the
bidimensional case which lends itself to experimental
confirmation. For each field $\psi$ (among density, velocity and pressure) in the cold region, we look for a solution of the form 
\beq \psi(\br,t) =
  \psi_0(\lambda) \left(1 + \delta\psi(\lambda)\, \cos(k \theta)
  \,t^s\right). \label{eq:perturbation} \eeq
with $\psi_0(\lambda)$ the unperturbed self-similar
profile, $\delta\psi(\lambda)$ the relative perturbation, and $s(k)$
the exponent of relative growth for a given angular
frequency $k$. This analysis is complicated by the fact that the underlying solution is neither uniform nor stationary.  We have to
resort to a method previously used in
conservative blasts~\cite{Kushnir05}: the exponent $s$ is used as
a free parameter, selected for each value of $k$ to minimize the difference between numerically integrated profiles and
theoretical values on the boundaries \cite{SuppMat}. We
thereby sample the dispersion relation $s(k)$: as seen in
Fig.~\ref{Fig:cartoon}, we predict the fastest growing
perturbation with $s\approx 0.3$. The value of the exponent and its independence on parameters $\alpha$ and $\phi_{rest}$ are both confirmed by simulations.

\paragraph*{Conclusion:}
Most extensions to the Taylor-von Neumann-Sedov blast exhibit either a breakdown of similarity, or 
self-similarity of the second kind, i.e. continuous dependence on dynamical parameters \cite{Bare96}. The dissipative blast studied here is exceptional in that its 
asymptotic expansion remains self-similar of the first kind, driven by inertial motion rather than fine-tuned by the dissipative processes themselves.  This property is generic to a large class of fluids:  regardless of its mechanism, bulk energy dissipation only comes into play to enforce the layered structure of the shock. Under weak conditions, the energy scales for coherent and incoherent motion decouple and each comes to dominate a different region: temperature resides in  a thin layer, pushed by a self-similarly  growing cold region.   
We have given a full hydrodynamic description of this structure, in excellent agreement with particle-level simulations, and thus brought to the fore the mechanisms 
underlying the similarity solution. Its cornerstone, the conservation of radial momentum per angular sector, stems from this decoupling of energy scales, reflected by the anisotropic pressure within the shell. Using the hydrodynamic profiles, we could perform a stability analysis, and successfully predict the existence and exponent of a
corrugation instability rooted in the cold region. These results invite further contact between kinetic and continuum approaches, and between fields, from plasmas to granular systems, to deepen our understanding of dissipative fluids.
\begin{acknowledgments}
We thank J.F Boudet, A. Vilquin and H. Kellay for insightful discussions.
\end{acknowledgments}


\begin{thebibliography}{99}
\bibitem{argument}
The radius $R$ of the blast depends on energy $E$, mass density (at rest) 
$\rho$ and time $t$. Their only dimensionless combination is $R^5 \rho / (E t^2)$, hence $R\propto t^{2/5}$, which readily generalizes to arbitrary dimension $d$ \cite{Bare96}.
\bibitem{note2}
If a solution grows self-similarly, spatial gradients are of order $1/R(t)$ and higher derivatives (as found in viscous or heat conduction-like terms) vanish faster.
\bibitem{rqene}The initial energy thus only sets the timescale for the subsequent evolution.
\bibitem{rque}
We used an Event-Driven algorithm. In order to avoid inelastic
collapse, collisions become elastic below a relative velocity threshold 
$\epsilon =10^{-10}$ \cite{Jabeen1,NiBC}.
\bibitem{rqtmp}
This granular temperature does not have a thermodynamic meaning.
\end{thebibliography}
\end{document}